\documentclass[10pt,aps,prb,twocolumn,superscriptaddress,showpacs]{revtex4-1}
\usepackage{graphicx}
\usepackage{amssymb}
\usepackage{amsmath}
\usepackage{appendix}
\usepackage{comment}
 \usepackage{color}

\begin{document}

\title{Theory of magnon-mediated tunnel magneto-Seebeck effect}

\author{Benedetta Flebus}
\affiliation{Institute for Theoretical Physics and Center for Extreme Matter and Emergent Phenomena, Utrecht University, Leuvenlaan 4, 3584 CE Utrecht, The Netherlands}
\affiliation{Department of Physics and Astronomy, University of California, Los Angeles, California 90095, USA}
\author{Gerrit E. W. Bauer}
\affiliation{Institute for Materials Research, Tohoku University, Sendai 980-8557, Japan}
\affiliation{Zernike Institute for Advanced Materials, University of Groningen, Nijenborgh 4, 9747 AG Groningen, The Netherlands}
\author{Rembert A. Duine}
\affiliation{Institute for Theoretical Physics and Center for Extreme Matter and Emergent Phenomena, Utrecht University, Leuvenlaan 4, 3584 CE Utrecht, The Netherlands}
\affiliation{Department of Applied Physics, Eindhoven University of Technology, PO Box 513,
5600 MB Eindhoven, The Netherlands}
\author{Yaroslav Tserkovnyak}
\affiliation{Department of Physics and Astronomy, University of California, Los Angeles, California 90095, USA}

\begin{abstract}
The tunnel magneto-Seebeck effect is the dependence of the thermopower of magnetic tunnel junctions on the magnetic configuration. It is conventionally interpreted in terms of a thermoelectric generalization of the tunnel magnetoresistance. Here, we investigate the heat-driven electron transport in these junctions associated with electron-magnon scattering, using stochastic Landau-Lifshitz phenomenology and quantum kinetic theory. Our findings challenge the widely accepted single-electron picture of the tunneling thermopower in magnetic junctions.
\end{abstract}
\maketitle

\section{Introduction}
\label{intro}

The interplay between spin, charge, and heat currents might provide new functionalities and increase the efficiency of existing thermoelectric technology. \cite{bauerSSC10,*bauerNATM12} The seminal work of Johnson and Silsbee\cite{johnsonPRB87} on nonequilibrium thermodynamics of spin-dependent transport foresaw the field that has since been dubbed \textit{spin caloritronics}. Only recently, however, the discovery of the spin Seebeck effect has revived general interest in the investigation of coupled heat and spin transport in metallic devices, which has since led to the observation of a number of striking phenomena. Among these, the tunneling magneto-Seebeck effect,\cite{walterNATM11,liebingPRL11,linNATC11} i.e., the dependence of the thermopower of a magnetic tunnel junction (MTJ) on the relative orientation of the two ferromagnetic layers, bridges spin caloritronics with conventional thermoelectrics, and offers  possibilities for a thermally-actuated magnetic data readout.\cite{boehnkeSR15}

The conventional interpretation of the tunnel magneto-Seebeck effect,\cite{walterNATM11,liebingPRL11,linNATC11} in terms of single-electron transport, rests on the assumption that the thermoelectric current is induced by the electron-hole asymmetry of the tunneling density of states at the Fermi level.\cite{CzernerPRB2011} A quantitative modeling of the experiments\cite{walterNATM11,liebingPRL11,linNATC11} by first-principles calculations\cite{CzernerPRB2011} did not lead to hard conclusions, however. On the experimental side (which is marred by a strong sample dependence), it is difficult, for instance, to determine the temperature drop over an ultrathin tunnel barrier, while calculations find a strong dependence on unknown details of disorder and alloy composition.

On another front, new experimental  evidence suggests that the magnon-drag  in the ferromagnetic bulk is playing a fundamental role in the thermopower: Watzman \textit{et al.}\cite{watzmanPRB16} observe a thermopower scaling as $T^{3/2}$ (and thus dominating over the single-electron diffusive contribution $\propto T$) over a broad temperature range in elemental transition metals, in agreement with the theoretically predicted magnon-drag contribution associated with the magnonic heat flux.\cite{lucassenAPL11,flebusPRB16} These findings call for a reassessment of the mechanism of the Seebeck effect in MTJs, where the relative importance of the electronic and magnonic contributions may be expected to parallel that in the bulk.\cite{Note1}

In this work, we report a theory of transport through a metallic ferromagnet (F)$|$insulator (I)$|$F junction subject to a thermal bias and evaluate the magnon-mediated contribution to the magneto-Seebeck effect in the semiclassical regime of magnetic fluctuations. In Sec.~\ref{scp}, we build upon the results of Ref.~[\onlinecite{tserkovPRB08tb}] to calculate the magnon-mediated magneto-Seebeck effect within the Landau-Lifshitz-Gilbert (LLG) stochastic phenomenology (assuming the adiabatic limit of the induced nonequilibrium electron transport). A more rigorous treatment is then developed in Sec.~\ref{qkt}, based on a quantum kinetic theory, which allows to systematically treat the coupled magnetic and electronic fluctuations (as needed for the determination of the full thermopower). It can handle diverse junctions and capture various microscopic mechanisms of the thermopower on equal footing, leading to the main results of this paper. In Sec.~\ref{bpp}, we offer an interpretation of the results of Secs.~\ref{scp} and \ref{qkt} in terms of the Berry phase, which links the semiclassical treatment based on the coherent ferromagnetic precession with the quantum approach that is centered on evaluating the electron-magnon scattering self-energies. The paper is closed with a discussion and outlook in Sec.~V.

\section{Stochastic charge pumping}
\label{scp}

The magnonic thermopower model in this section is based on the charge-pumping concept by coherent magnetic dynamics in magnetic tunnel junctions.\cite{tserkovPRB08tb,xiaoPRB08cp} After a brief review, we generalize this scheme to model the charge current in temperature-biased MTJs using the stochastic LLG equation.

\subsection{Phenomenology of magnetic pumping}

\begin{figure}[pt]
\includegraphics[width=0.7\linewidth]{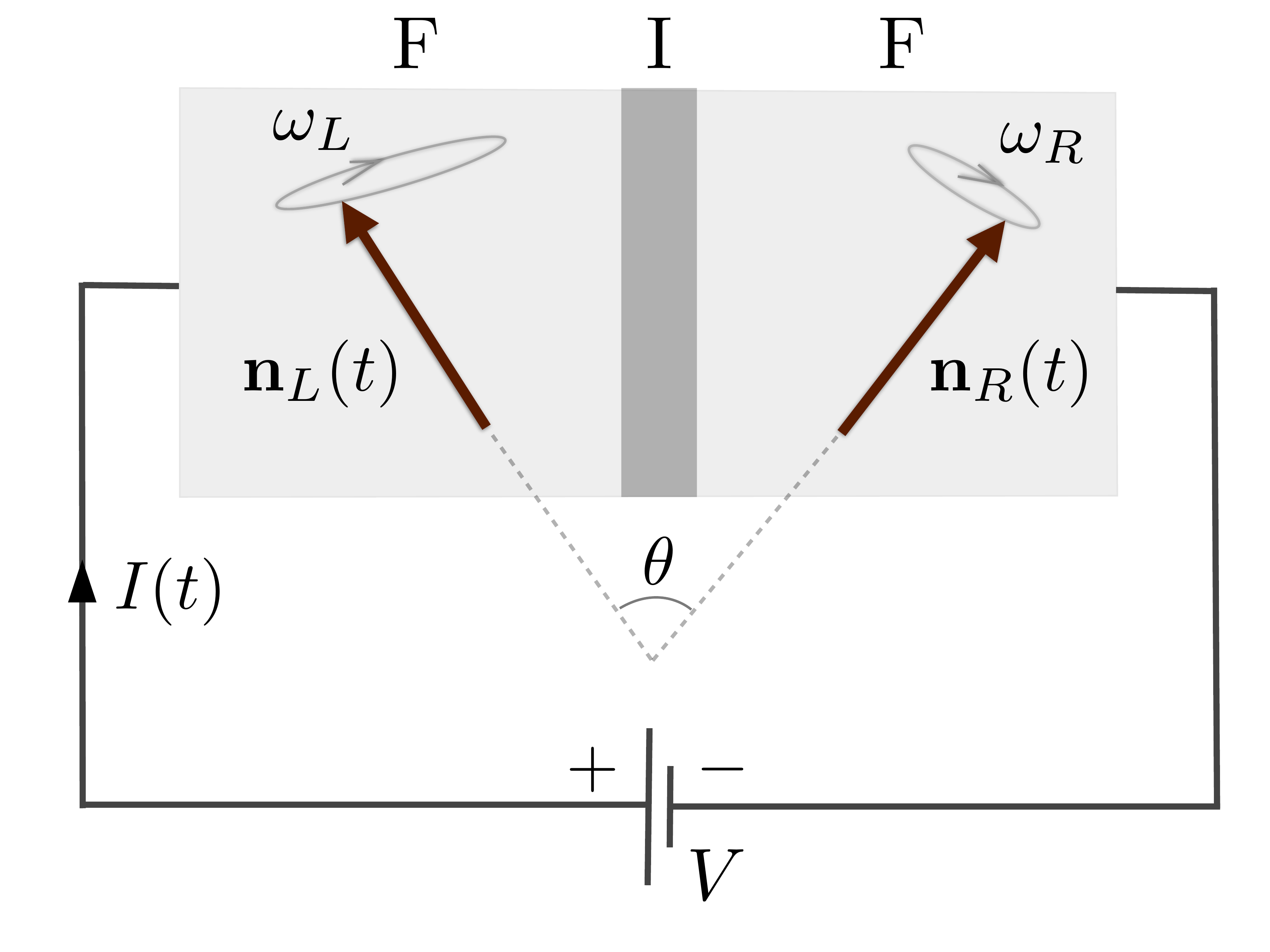}
\caption{An F$|$I$|$F junction circuit, subjected to an applied voltage and coherent microwave dynamics of the monodomain order parameters, $\mathbf{n}_L(t)$ and $\mathbf{n}_R(t)$. The ensuing time-dependent electric current, Eq.~\eqref{equation1}, consists of an Ohmic, $\propto G$, and adiabatically pumped, $\propto\eta$, contributions.}
\label{figurecircuit}
\end{figure}

We consider the simplest case of a structurally mirror-symmetric 
junction of two metallic isotropic monodomain ferromagnets separated by a thin tunneling barrier, as shown in Fig.~\ref{figurecircuit}. At low temperatures, i.e., $T \ll T_{C},T_{F}$ (with $T_{C}$ and $T_{F}$ being the Curie and Fermi temperature, respectively), the electric current is controlled by the voltage $V$ applied to the junction and the (unit-vector) spin-density order parameter dynamics of the left and right magnetic layer, $\mathbf{n}_L(t)$ and $\mathbf{n}_R(t)$.\cite{Note2} Focusing on the adiabatic limit of the magnetic charge pumping, the electric current can be written in terms of the instantaneous $\mathbf{n}_i(t)$ and $\partial_t\mathbf{n}_i(t)$ (with $i=R,L$) as~\cite{Note3}
\begin{equation}
I=GV+\eta\left(\mathbf{n}_{L}\cdot\mathbf{n}_{R}\times\partial_{t}\mathbf{n}_{R}-\mathbf{n}_{R}\cdot\mathbf{n}_{L}\times\partial_{t}\mathbf{n}_{L}\right)\,.
\label{equation1}
\end{equation}
Here, $G \equiv G(\mathbf{n}_L\cdot\mathbf{n}_R)$ is the electric conductance of the junction, that, assuming isotropicity in spin space, depends solely on $\mathbf{n}_{L} \cdot \mathbf{n}_{R}=\cos\theta$. The parameter $\eta$ depends on the spin-dependent electronic structure of the system,\cite{xiaoPRB08cp,tserkovPRB08tb} as discussed below. For simplicity (and as justified below in our tunneling model), we disregard the possible dependence of $\eta$ on $\theta$. In the absence of applied voltage, Eq.~(\ref{equation1}) gives, for a steady (left-hand) precession of $\mathbf{n}_L$ at  frequency $\omega$ with cone angle $\theta$ around a static $\mathbf{n}_R$, the magnetically pumped current
\begin{equation}
I_m=\eta\omega\sin^2\theta\,.
\label{equation2}
\end{equation}

\subsection{Tunneling Hamiltonian}
Our Hamiltonian for conduction electrons in a symmetric F$|$I$|$F junction reads
\begin{equation}
\mathcal{H}=\mathcal{H}_L+\mathcal{H}_R+\mathcal{H}'\,,
\label{equation3}
\end{equation}
where $\mathcal{H}_{L,R}$ are the bulk Hamiltonians for the respective magnetic leads and $\mathcal{H}'$ is the tunneling Hamiltonian. The  Hamiltonian of the ferromagnetic leads can be written  as (with $i=L,R$)
\begin{align}
\mathcal{H}_{i}=&\sum_{\mathbf{k},\sigma=\uparrow, \downarrow}\varepsilon_{\mathbf{k}}c^{\dagger}_{\mathbf{k},\sigma} c_{ \mathbf{k}, \sigma}\nonumber\\
&-\frac{\Delta}{2}\sum_{\sigma\sigma'}\int d^3r\,\psi_{\sigma}^\dagger(\mathbf{r})\mathbf{n}_{i}(\mathbf{r},t)\cdot\hat{\boldsymbol{\sigma}}_{\sigma\sigma'}\psi_{\sigma'}(\mathbf{r})\,,
\label{equat4}
\end{align}
in a mixed momentum-position representation.
Here, the 
dispersion $\varepsilon_{\mathbf{k}}$ captures the (nonmagnetic) 
band structure, $\hat{\boldsymbol{\sigma}}$ is the vector of the Pauli 
matrices, $c_{\mathbf{k}\sigma}$ and $\psi_{\sigma}(\mathbf{r})=\sum_{\mathbf{k}} c_{\mathbf{k},\sigma} e^{i \mathbf{k} r}/\sqrt{\mathcal{V}}$ (with $\mathcal{V}$ being the volume of both ferromagnets)  are the electron field operators (for the $i$th side, having omitted the corresponding label) obeying the fermionic commutation relations:  $\{ c_{\mathbf{k},\sigma}, c^{\dagger}_{\mathbf{k'},\sigma'}\}=\delta_{\mathbf{k} \mathbf{k'}} \delta_{\sigma\sigma'}$ and $\{\psi_{\sigma}(\mathbf{r}),\psi_{\sigma'}^{\dagger}(\mathbf{r}') \}=\delta (\mathbf{r} - \mathbf{r}') \delta_{\sigma\sigma'}$.  The interaction between the itinerant electrons and the spin density  order parameter $\mathbf{n}_{i}$ is parametrized by a uniform exchange splitting $\Delta$. The  tunneling of electrons through the insulating barrier described by the Hamiltonian
\begin{equation}
\mathcal{H}'=\sqrt{\frac{\hbar}{2\pi}}\frac{\tau}{\mathcal{V}}\sum_{\mathbf{k},\mathbf{k}',\sigma}c_{\mathbf{k},\sigma,L}^\dagger c_{\mathbf{k}',\sigma,R} + {\rm H.c.}
\label{Hprime}
\end{equation}
is assumed to preserve spin, to not depend on $\mathbf{n}$, and is ideally 
diffuse, i.e., 
 the tunneling matrix elements $\tau$ do not depend on spin and momentum indices.  While such spin- and momentum-independent tunneling presents a rather simplistic view on the problem (see, e.g., Ref.~[\onlinecite{slonczewskiPRB05}] for a more thorough analysis) and, e.g., does not include the symmetry-based selection rules that cause large magnetoresistance in epitaxial tunneling barriers such as MgO, it should capture the essence of the collective effects that are of interest to us. In particular, it provides us with a simple model to assess the magnonic contribution relative to the purely electronic one.
 
The tunneling Hamiltonian (\ref{Hprime}) was used in Ref.~[\onlinecite{tserkovPRB08tb}] to calculate (through the rotating-frame approach) the parameter $\eta$ in Eq.~\eqref{equation2}, yielding
\begin{equation}
\eta=-\frac{e\hbar}{4}|\tau|^{2}(D^{2}_{\uparrow} - D^{2}_{\downarrow})=-e\hbar|\tau|^2\bar{D}^2P\,.
\label{equation6}
\end{equation}
Here, $e$ is the carrier charge ($<0$, for electrons) and $D_{\sigma}$ is the spin-$\sigma$ (along $\mathbf{n}$) density of states (per unit volume $\mathcal{V}$) in the magnetic leads, $\bar{D}\equiv(D_\uparrow+D_\downarrow)/2$ and $P\equiv(D_\uparrow-D_\downarrow)/(D_\uparrow+D_\downarrow)$. The conductance (neglecting magnon-assisted terms) is
\begin{equation}
G=2e^2|\tau|^2\bar{D}^2(1+P^2\cos\theta)=-\eta\frac{2e}{\hbar}\frac{1+P^2\cos\theta}{P}\,,
\end{equation}
so that the voltage induced by the magnetization dynamics in an open circuit (i.e., with zero current) is
\begin{equation}
V|_{I=0}=-\frac{\eta}{G}\omega\sin^2\theta=\frac{\hbar\omega}{2e}\frac{P\sin^2\theta}{1+P^2\cos\theta}\,.
\end{equation}
In the following, we are interested in analogous voltages induced by thermally-induced magnetic fluctuations.

\subsection{LLG theory of thermopower}
\label{LLGth}

\begin{figure}[pt]
\includegraphics[width=0.7\linewidth]{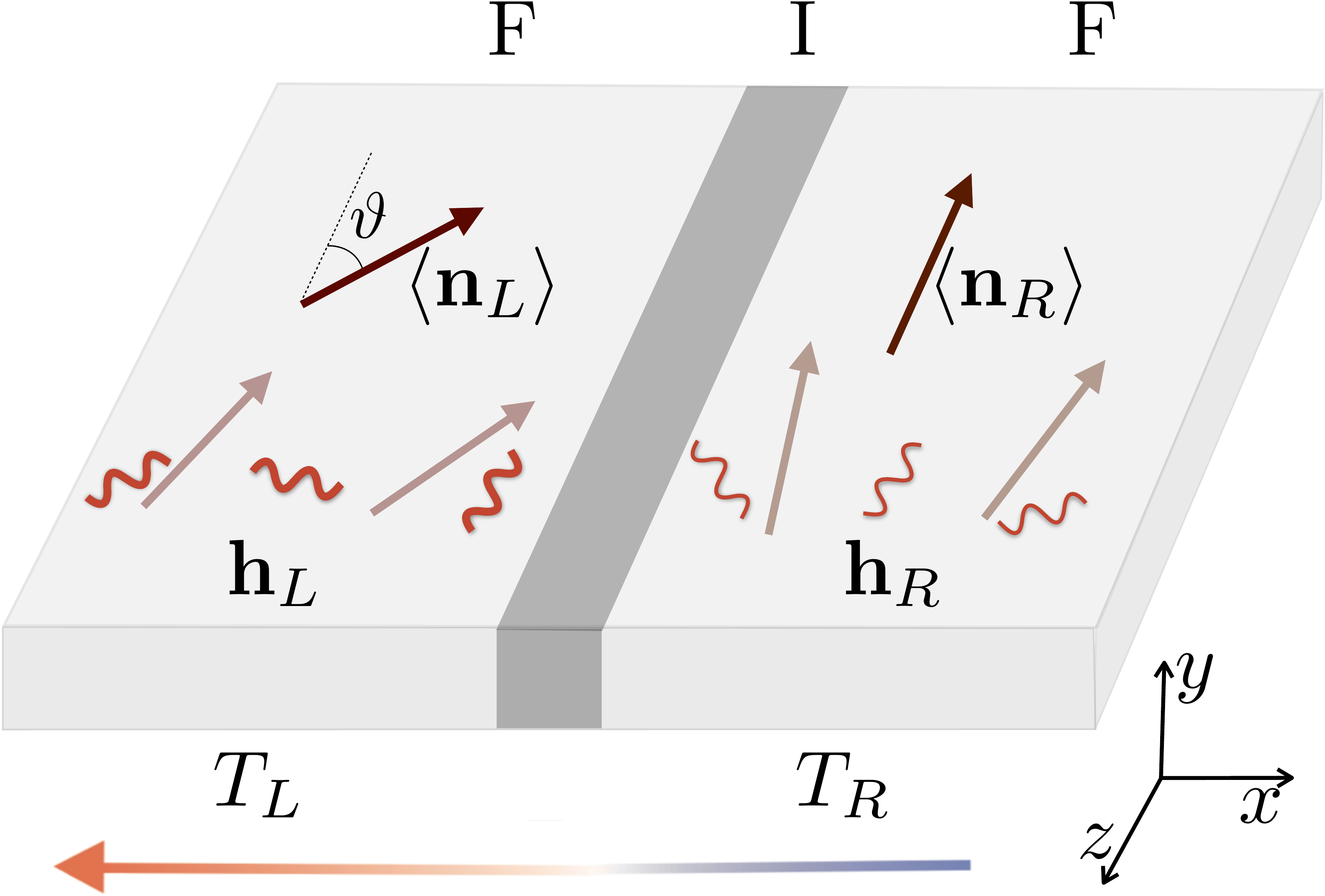}
\caption{Schematic of a thermally-biased F$|$I$|$F junction. Stochastic fluctuations around the equilibrium orientations of the spin-density order parameters, $\langle\mathbf{n}_{R}\rangle\parallel-\mathbf{z}$ and $\langle\mathbf{n}_{L}\rangle$ (tilted by an angle $\vartheta$), are induced by the Langevin fields $\mathbf{h}_R$ and $\mathbf{h}_L$ (wiggly lines), whose intensity is greater in the left lead when its temperature $T_L>T_R$.}
\label{figurestochastic}
\end{figure}

Applying Eq.~\eqref{equation1} to a thermally-biased MTJ, as in Fig.~\ref{figurestochastic}, the short-circuit (i.e., zero-voltage) current, due to the magnetic pumping, is given by
\begin{equation}
I_m=\eta  \langle  \mathbf{n}_{L} \cdot  \mathbf{n}_{R} \times \partial_{t} \mathbf{n}_{R}   -  \mathbf{n}_{R} \cdot \mathbf{n}_{L} \times \partial_{t} \mathbf{n}_{L} \rangle\,.
\label{equation7}
\end{equation}
The averaging $\langle\dots\rangle$ is carried out here over the steady-state stochastic fluctuations of the spin density orientations, $\mathbf{n}_{L,R}(\mathbf{r},t)$. For noise sources that are uncorrelated across the junction (as is the case for the local magnetization damping), the averaging in Eq.~\eqref{equation7} can be correspondingly factored out:
\begin{align}
\langle\mathbf{n}_{L}\cdot\mathbf{n}_{R}\times\partial_{t}\mathbf{n}_{R}\rangle&=\langle\mathbf{n}_{L}\rangle\cdot\langle\mathbf{n}_{R}\times\partial_{t}\mathbf{n}_{R}\rangle\nonumber\\
&\approx-\cos\vartheta\,\mathbf{z}\cdot\langle\mathbf{n}_{R}\times\partial_{t}\mathbf{n}_{R}\rangle\,,
\end{align}
and similarly for the second term in Eq.~\eqref{equation7}. Here, we are assuming small-angle fluctuations, which govern the thermopower to the leading order in $T/T_C$.  $\vartheta$ is the angle between the equilibrium values of the spin-order parameters $\langle\mathbf{n}_{L}\rangle$ and $\langle\mathbf{n}_{R}\rangle$, and we used $\langle\mathbf{n}_{R}\times\partial_{t}\mathbf{n}_{R}\rangle\propto\langle\mathbf{n}_{R}\rangle\propto\mathbf{z}$. The problem now reduces to evaluating $f(T)\equiv\mathbf{z}\cdot\langle\mathbf{n}\times\partial_{t}\mathbf{n}\rangle$ in a uniform bulk magnet at temperature $T$, with $\langle\mathbf{n}\rangle\parallel-\mathbf{z}$. The thermally-pumped current is then given by
\begin{equation}
I_m=2\eta\cos\vartheta[f(T_L)-f(T_R)]\,,
\end{equation}
where the factor of 2 stems from the Neumann (exchange) boundary condition for the magnetic fluctuations at the junction, which doubles the power of thermal noise and the associated pumping.\cite{hoffmanPRB13,kapelrudPRL13}

The dynamics of the order parameter $\mathbf{n}$  is governed by the stochastic LLG equation~\cite{LLGstochastic}
\begin{equation}
s ( 1 + \alpha \mathbf{n} \times ) \partial_{t} \mathbf{n} + \mathbf{n} \times (H\mathbf{z} + \mathbf{h}) + \sum_{i} \partial_{i} \mathbf{j}_{s,i}=0\,.
\label{equation9}
\end{equation}
where $s$ is the saturation spin density, $\alpha$  the dimensionless Gilbert damping,  $\mathbf{j}_{s,i}\equiv - A \mathbf{n} \times \partial_{i} \mathbf{n}$  the magnetic spin-current density, which is proportional to the exchange stiffness $A$, and $H$ denotes a magnetic field oriented along the $z$ axis. Furthermore, $\mathbf{h}$ is the Langevin field conjugated to the Gilbert damping, with the correlator
\begin{equation}
\langle h_{i}(\mathbf{r}, \omega) h^{*}_{j}(\mathbf{r}', \omega') \rangle = \frac{2\pi \alpha s \hbar \omega \delta_{ij} \delta (\mathbf{r}- \mathbf{r}') \delta (\omega - \omega')}{\tanh \frac{\hbar \omega}{2 k_{B} T(\mathbf{r})}}\,.
\label{equation10}
\end{equation}
Following the procedure of Ref.~[\onlinecite{flebusPRB16}],  we solve Eq.~(\ref{equation9}) to leading order in  the transverse spin-density fluctuations, and arrive at (in the limit $\alpha\ll1$) 
\begin{equation}
I_m= \eta\frac{2\cos\vartheta}{s}  \int \frac{d^{3} \mathbf{q} }{(2\pi)^{3}} \; \hbar \omega_{\mathbf{q}} \left[  \coth\frac{\hbar \omega_{\mathbf{q}}}{2 k_{B} T_{L}} - \coth\frac{\hbar \omega_{\mathbf{q}}}{2 k_{B} T_{R}} \right]\,,
\label{equation11}
\end{equation}
where $\omega_{\mathbf{q}}$ is the magnon dispersion. For a small temperature bias, $\delta T=T_L-T_R$, and in the regime  $\hbar H/s\ll k_BT$, where  $\omega_{\mathbf{q}}\approx A q^{2}/s$,   Eq.~(\ref{equation11}) boils down to
\begin{equation}
I_m=\eta\frac{2Jk_B}{\pi^{2} \hbar}\cos\vartheta\left( \frac{T}{T_c} \right)^{3/2}\delta T\,.
\label{equation12}
\end{equation}
Here,
\begin{equation}
J \equiv \int_{0}^{\infty} dx \frac{\sqrt{2} x^{6}}{\sinh^{2} x^{2}}  \sim 1
\end{equation}
is a numerical constant and $k_B T_c\equiv A(\hbar/s)^{1/3}$ approximates the Curie temperature. The temperature exponent $3/2$ coincides with that of the related bulk magnon-drag thermopower.\cite{lucassenAPL11,flebusPRB16,watzmanPRB16} The magnetic junction-mediated Seebeck coefficient is thus found to be
\begin{equation}
\mathcal{S}_m=-\left.\frac{V}{\delta T}\right|_{I_m=0}=-\frac{P\cos\vartheta}{1+P^2\cos\vartheta}\frac{Jk_B}{\pi^{2}e}\left( \frac{T}{T_c} \right)^{3/2}\,,
\label{Sm}
\end{equation}
where the full current (neglecting for now the free-electron thermopower) is $I=GV+I_m$.

We can gain additional insights by rewriting Eq.~(\ref{equation11}) as an integral over the magnon energy $\epsilon=\hbar\omega$: 
\begin{equation}
I_m=\eta\frac{4\cos\vartheta}{s} \int d \epsilon D(\epsilon) \epsilon  \left[n_{B}(\beta_L\epsilon) -n_{B}(\beta_R\epsilon)   \right]\,.
\label{equation14}
\end{equation}
Here, $n_{B}(x)\equiv(e^{x}-1)^{-1}$ is the Bose-Einstein distribution function, $\beta\equiv1/k_{B} T$ is the inverse temperature, and $D(\epsilon)$ is the magnon density of states (per unit volume). Eq.~(\ref{equation14}) shows that the pumped charge current is proportional to the difference in the magnon energy density in the two magnetic leads. The contribution associated with a single magnon mode at energy $\epsilon$ (say, in the left ferromagnet), for $\vartheta\to0$, is thus $I_{m}=4\eta\omega\,n_{B}/S$, where
\begin{equation}
S\equiv s\mathcal{V}/\hbar
\end{equation}
is the dimensionless macrospin. Relating this to the (small) precession-cone angle $\theta$ according to $n_{B}=(1-\cos\theta)S\approx\theta^2S/2$ gives $I_{m}\approx2\eta\omega\theta^2$, which reproduces Eq.~\eqref{equation2} (times the aforementioned factor of 2 due to the Neumann boundary condition).

The semiclassical formalism, based on magnetic charge pumping, is, however, not complete. For example, if the ferromagnets are subjected to different magnetic fields $H_{L(R)}$ (on the left and right, respectively), evaluating Eq.~\eqref{equation7} according to the above procedure leads to
\begin{align}
I_m= \eta\frac{4\cos\vartheta}{s} \int d \epsilon &D(\epsilon) \big[(\epsilon + \epsilon_L) n_{B}(\beta_L(\epsilon + \epsilon_L) )  \nonumber \\
&-(\epsilon + \epsilon_R) n_{B}(\beta_R(\epsilon + \epsilon_R) )  \big]\,,
\label{equation19new}
\end{align}
where $\epsilon_{L(R)}\equiv\hbar H_{L(R)}/s$ are the magnon gaps, and the energy integration has been shifted to start at the respective magnon-band edges. Equation (\ref{equation19new}) would thus imply that a finite current $I$ is possible even when $T_L=T_R$, if $H_L\neq H_R$. This unphysical result stems from disregarding possible additional contributions to the thermopower, which are rooted in the electronic spin-current noise. In particular, the transverse spin-current noise\cite{forosPRL05} may be rectified by the magnetic fluctuations that it triggers, analogously to the rectification of the spin current underlying the spin Seebeck effect.\cite{hoffmanPRB13,xiaoPRB10} In effect, the stochastic LLG treatment includes thermal noise of magnetic dynamics but does not consider the electronic noise sources. To remedy this, we now turn to a more systematic field-theoretic treatment that captures magnetic and electronic fluctuations on equal footing, accounting consistently for quantum as well as thermal fluctuations. In Sec.~\ref{bpp}, we will identify the deficiencies of the stochastic Langevin theory from the field-theoretic perspective.

\section{Quantum-kinetic theory}
\label{qkt}
In this section, we introduce the second-quantized Hamiltonian and derive its spectral functions. We deploy the latter to calculate the tunneling current for parallel and antiparallel configurations, followed by a generalization to noncollinear magnetization configurations and the presence of spin-flip scattering in the junction.

\subsection{Spin-dependent spectral functions}

Thermally-induced electron tunneling in magnetic junctions can be treated systematically by a quantum-kinetic formalism,  a general framework to address  transport through weak links. We start by quantizing the magnetic orientation $\mathbf{n}=(n_x,n_y,n_z)$  in Eq.~\eqref{equat4} by the Holstein-Primakoff transformation to leading order in small angle fluctuations:\cite{holsteinPR40}
\begin{equation}
n_x-in_y(\mathbf{r})\approx\sqrt{\frac{2\hbar}{s}} \hat{\phi}^{\dagger}(\mathbf{r}), \; \; n_{z}(\mathbf{r}) = \frac{\hbar}{s}\hat{\phi}^{\dagger}(\mathbf{r}) \hat{\phi}(\mathbf{r})-1\,,
\label{eq20}
\end{equation}
where the magnon field operator $\hat{\phi}(\mathbf{r})=\sum_{\mathbf{k}} a_{\mathbf{k}} e^{i \mathbf{k}\cdot\mathbf{r}}/\sqrt{\mathcal{V}}$ obeys the bosonic commutation relation $[\hat{\phi}(\mathbf{r}), \hat{\phi}^{\dagger}(\mathbf{r})]=\delta (\mathbf{r}-\mathbf{r}')$, and $a_\mathbf{k}$ ($a^\dagger_\mathbf{k}$) is the magnon creation (annihilation) operator, with $[a_\mathbf{k},a^\dagger_{\mathbf{k}'}]=\delta_{\mathbf{k}\mathbf{k}'}$. We first address  the parallel alignment of the magnetizations, $\vartheta=0$. 

Substituting Eqs.~(\ref{eq20}) into Eq.~(\ref{equat4}) and adding the free-magnon energy leads to second-quantized  Hamiltonian for a bulk metallic magnet:
\begin{align}
\mathcal{H}_{L(R)}=&\sum_{ \mathbf{k}, \sigma} \varepsilon_{\mathbf{k},\sigma} c^{\dagger}_{\mathbf{k},\sigma} c_{ \mathbf{k}, \sigma}+ \sum_{\mathbf{q}} \epsilon_{\mathbf{q}} a^{\dagger}_{\mathbf{q}} a_{\mathbf{q}}\nonumber\\
&- \frac{\Delta}{\sqrt{2S}}\sum_{ \mathbf{k}, \mathbf{q}}\left(c^{\dagger}_{ \mathbf{k}+\mathbf{q}, \uparrow} c_{ \mathbf{k}, \downarrow} a_{\mathbf{q}} + \text{H.c.} \right)\nonumber\\
& - \frac{\Delta}{2S}\sum_{ \mathbf{k},\mathbf{q}, \boldsymbol{\kappa}, \sigma} \sigma\,c^{\dagger}_{\mathbf{k}+\boldsymbol{\kappa},\sigma} c_{ \mathbf{k}, \sigma}  a^\dagger_{\mathbf{q}-\boldsymbol{\kappa}}a_{\mathbf{q}}\,,
\label{equation17}
\end{align}
where $\varepsilon_{\mathbf{k},\sigma}=\varepsilon_{\mathbf{k}} + \sigma \Delta/2$ is the (mean-field) spin-dependent electron energy, with $\sigma=\pm 1$ corresponding respectively to spin $\uparrow,\downarrow$ along the $z$ axis, and $\epsilon_{\mathbf{q}}= \hbar \omega + A q^{2}/s$ is the  magnon dispersion relation. The tunneling Hamiltonian~(\ref{Hprime}) connects the two magnetic leads, each described by Eq.~(\ref{equation17}). The magnon-electron interaction affects the electronic transport by dressing the electronic states and by allowing for inelastic scattering processes.
Here we focus on the electron interaction with a single magnon mode, at frequency $\omega$, disregarding (higher-order) multimagnon processes. The corresponding (retarded) electron self-energies read as~\cite{appelbaumPR69,*woolseyPRB70,*macdonaldPRL98}
\begin{align}
\Sigma_{\mathbf{k},\downarrow}(\varepsilon) &= \frac{\Delta^2}{2 S} \frac{n_{F}(\beta \varepsilon_{\mathbf{k},\uparrow}) + n_{B}(\beta \hbar \omega)}{\varepsilon + \hbar \omega - \varepsilon_{\mathbf{k}, \uparrow}+i0^+}\,, \nonumber \\
\Sigma_{\mathbf{k},\uparrow}(\varepsilon) &= \frac{\Delta^2}{2 S} \frac{n_{F}(-\beta \varepsilon_{\mathbf{k} \downarrow}) + n_{B}(\beta \hbar \omega)}{\varepsilon - \hbar \omega - \varepsilon_{\mathbf{k}, \downarrow}+i0^+}\,.
\end{align}
Here, on the right-hand sides, we implicitly include the (forward-scattering) correction $-\sigma\Delta\,n_B(\beta\hbar\omega)/2S$ to $\varepsilon_{\mathbf{k},\sigma}$, which arises from the last term in Eq.~\eqref{equation17}.
 $n_{F}(x)\equiv(e^{x}+1)^{-1}$ is the Fermi-Dirac distribution function and $0^+$ is an infinitesimal positive.
To linear order in  $\hbar \omega / \Delta, n_{B}/S, n_{F}(\beta \varepsilon_{\mathbf{k}, \uparrow})/S, [1 - n_{F}(\beta \varepsilon_{\mathbf{k} \downarrow})]/S \ll 1$, the  spin-down spectral function $A_{\mathbf{k},\downarrow}(\varepsilon)$ has two peaks (with, henceforth, bare energies $\epsilon_{\mathbf{k}, \sigma}$) at
\begin{align}
\tilde{\varepsilon}_{\mathbf{k}, \sigma} =\varepsilon_{\mathbf{k},\sigma} &- \hbar \omega \left[ \delta_{\sigma,+} - \sigma \frac{n_{B}(\beta \hbar \omega)}{2S} \right] \nonumber\\
&+ \sigma (\Delta + \hbar \omega) \frac{n_{F}(\beta \epsilon_{\mathbf{k},\uparrow})}{2S} \,,
\label{epsilontilde}
\end{align}
with the respective spectral weights $\delta_{\sigma,-} +\sigma [ n_{B}(\beta \hbar \omega) + n_{F}(\varepsilon_{\mathbf{k},\uparrow})]/2S$. Hence,
\begin{align}
\frac{A_{\mathbf{k},\downarrow}(\varepsilon)}{2\pi}=&\left[ 1 - \frac{n_{B}(\beta \hbar \omega)+ n_{F}(\beta \varepsilon_{\mathbf{k} ,\uparrow})}{2S} \right] \delta (\varepsilon - \tilde{\varepsilon}_{\mathbf{k},\downarrow}) \nonumber \\
& + \frac{n_{B}(\beta \hbar \omega) + n_{F}(\beta \varepsilon_{\mathbf{k},\uparrow})}{2S} \delta (\varepsilon - \tilde{\varepsilon}_{\mathbf{k}, \uparrow})\,.
\label{Adown}
\end{align}
Similar considerations for the spin-up spectral function $A_{\mathbf{k},\uparrow}(\varepsilon)$ lead us to
\begin{align}
\frac{A_{\mathbf{k},\uparrow}(\varepsilon)}{2\pi}=& \left[ 1 -\frac{n_{B}(\beta \hbar \omega) + n_{F}(-\beta \varepsilon_{\mathbf{k}, \downarrow})}{2S} \right] \delta ( \varepsilon - \bar{\varepsilon}_{ \mathbf{k}, \uparrow}) \nonumber \\
&+ \frac{n_{B}(\beta \hbar \omega) + n_{F}(-\beta \varepsilon_{\mathbf{k}, \downarrow})}{2S} \delta ( \varepsilon - \bar{\varepsilon}_{\mathbf{k}, \downarrow})\,,
\label{Aup}
\end{align}
with peaks at
\begin{align}
\bar{\varepsilon}_{\mathbf{k}, \sigma} =\varepsilon_{\mathbf{k}, \sigma} &+ \hbar  \omega\left[ \delta_{\sigma, -} + \sigma \frac{n_{B}(\beta \hbar \omega)}{2S} \right] \nonumber \\
& - \sigma (\Delta + \hbar \omega) \frac{n_{F}(-\beta \varepsilon_{\mathbf{k}, \downarrow})}{2S} \,.
\label{equazione26}
\end{align}
The spectral functions reflect the lowest-order electron-magnon scattering, i.e., a spin-up electron flipping spin by emitting a magnon and the inverse process, both introduced by the exchange interaction term in the second line of Eq.~(\ref{equation17}).

\subsection{Magnon-assisted electron tunneling}

The expression for the tunneling current following from the Hamiltonian (\ref{Hprime}) reads~\cite{bruusBOOK04}
\begin{align}
I=\frac{e |\tau|^{2}}{2\pi \mathcal{V}^{2}}\sum_{\mathbf{k}, \mathbf{k}', \sigma}\int_{-\infty}^{\infty}  \frac{d\varepsilon}{2\pi}&\left[n_{F}(\beta_L\varepsilon)- n_{F}(\beta_R\varepsilon)\right] \nonumber \\
&\times A_{\mathbf{k},\sigma,L}(\varepsilon)A_{\mathbf{k}',\sigma,R} (\varepsilon)\,.
\label{21}
\end{align}
The identities
\begin{equation}
n_{F}(x_{1}) n_{F}(-x_{2}) \equiv n_{B} (x_{1} - x_{2} ) [n_{F}(x_{2}) - n_{F}(x_{1}) ]\,,
\end{equation}
and
\begin{equation}
\int_{-\infty}^\infty dx [ n_{F}(x) - n_{F}(x + x_{1}) ] \equiv x_{1}\,,
\end{equation}
are useful to decompose the net tunneling current $I$, Eq.~\eqref{21}, into the free-electron, $I_f$, and the magnon-assisted contributions.  At low temperatures, we can approximate the spin-$\sigma$ (along $\mathbf{n}$) density of states $D_{\sigma}$ by its value at the Fermi level, i.e., $D_{\sigma}(\varepsilon) \cong D_{\sigma}$, within the thermal window set by $k_BT$. By this approximation, we omit the (Mott) free-electron thermoelectric current, $I_f\propto T /T_F$, and magnonic corrections $\mathcal{O}(T/T_F)$.

The leading-order magnonic contribution $I^{(P)}$ to the current, in the parallel (P) configuration, then reads
\begin{align}
I^{(P)}=&\frac{ e |\tau|^{2}}{S}D_{\uparrow} D_{\downarrow}\int d \varepsilon\left[n_{F}(-\beta_R\varepsilon)n_{F} \left(\beta_L(\epsilon + \hbar \omega) \right)\right.\nonumber\\
&\hspace{15mm}\left.-n_{F}(-\beta_L\varepsilon) n_{F}\left( \beta_R(\epsilon + \hbar \omega)\right)\right]\,.
\label{Equation29}
 \end{align}
For a small temperature bias $\delta T= T_{L} -T_{R}$, we can express Eq.~(\ref{Equation29}) in terms of a dimensionless integral:
\begin{align}
I^{(P)}=& - \frac{  e |\tau|^{2}}{2 S}D_{\uparrow} D_{\downarrow} k_B\delta T\nonumber \\
&\times \int \frac{dx\,x}{\cosh^{2} x} \left[ \frac{e^{-(y+x)}}{\cosh (y+x)} + \frac{e^{-(y-x)}}{\cosh(y-x)} \right]\,,
\label{equazione31}
\end{align}
where  $x=\beta\epsilon/2$ and $y=\beta\hbar \omega/2$. Making use of the  identity
\begin{align}
\int\frac{dx\,x}{\cosh^{2} x} \frac{e^{-(y \pm x)}}{\cosh(y \pm x)} \equiv \pm \frac{y^{2}}{\sinh^{2} y}\,,
\end{align}
Equation~(\ref{equazione31}) vanishes, in agreement with Ref.~[\onlinecite{mccannPRB02}].

For the antiparallel (AP) alignment of the magnetizations, i.e., $\vartheta\to\pi$, the magnonic contribution to the current becomes $I^{(AP)}= I' + I''$, with 
\begin{align}
I'&= \frac{ e |\tau|^{2}}{2 S}  (D^{2}_{\uparrow} + D^{2}_{\downarrow} )\int d \varepsilon  \left[n_{F}(-\beta_R\varepsilon) n_{F}\left(\beta_L(\epsilon + \hbar \omega)\right)\right. \nonumber \\
&\hspace{25mm} \left.-n_{F}(-\beta_L\varepsilon) n_{F}\left( \beta_R(\epsilon + \hbar \omega) \right) \right]\,,
\label{EQUZ31} \\
I''&=\frac{e |\tau|^{2}}{2 S} (D^{2}_{\uparrow} - D^{2}_{\downarrow}) \hbar \omega[n_{B}(\beta_L\hbar \omega) - n_{B}(\beta_R\hbar \omega)]\,.
\label{eq22}
\end{align}
Similarly to Eq.~(\ref{Equation29}), Eq.~(\ref{EQUZ31}) vanishes in linear response.
In terms of the parameter $\eta$ (Eq.~\ref{equation6}), on the other hand,  Eq.~(\ref{eq22}) becomes
\begin{align}
I''=&-\eta\frac{2}{s \mathcal{V} } \hbar \omega  [n_{B}(\beta_L\hbar \omega) - n_{B}(\beta_R\hbar \omega)]\,.
\label{eq23}
\end{align}
In order to account for the magnetic Neumann boundary conditions, which enhance the power of finite-wavelength magnetic fluctuations at the interface, we should multiply this result by two.\cite{hoffmanPRB13,kapelrudPRL13} Integrating, furthermore, over the continuum of the magnon modes, we finally get
\begin{equation}
I=-\eta\frac{4\sin^2\frac{\vartheta}{2}}{s} \int d \epsilon D(\epsilon) \epsilon  \left[n_{B}(\beta_L\epsilon) -n_{B}(\beta_R\epsilon) \right]\,.
\label{Iqk}
\end{equation}
Here, we restored an arbitrary alignment by interpolating between the parallel and antiparallel cases (according to the $\cos\frac{\vartheta}{2}$ and $\sin\frac{\vartheta}{2}$ spin-dependent tunneling matrix elements, which enter into the generalization of Eq.~\eqref{21} to arbitrary $\vartheta$). While Eqs.~\eqref{equation14} and \eqref{Iqk} agree regarding the overall strength of the magnon-assisted current (and coincide exactly in the antiparallel case), the angular dependence differs. Most importantly, Eq.~\eqref{Iqk} vanishes in the P case (i.e., for $\vartheta\to0$), while Eq.~\eqref{equation14} reaches maximum. We will discuss the origin of this discrepancy in Sec.~\ref{bpp}.

Our final expression for the linear-response thermopower is obtained analogously to Eq.~\eqref{Sm}:
\begin{equation}
\mathcal{S}=\frac{P\sin^2\frac{\vartheta}{2}}{1+P^2\cos\vartheta}\frac{Jk_B}{\pi^{2}e}\left( \frac{T}{T_c} \right)^{3/2}\,.
\label{Sfinal}
\end{equation}
This is a central result of this paper. For the electron-like carriers (i.e., $e<0$) with ferromagnetic alignment between itinerant spin density and the order parameter $\mathbf{n}$ (i.e., $P>0$), the thermopower is negative, as in a simple metal. Equation \eqref{Sfinal} is consistent with the results of Ref.~\onlinecite{mccannPRB02} in the collinear limits (i.e., $\vartheta=0$ or $\pi$).

\subsection{Electron spin flips}

The present formalism can be extended to include the heretofore disregarded electron spin flips in the tunneling Hamiltonian. These can be caused by elastic electron-impurity or inelastic electron-phonon scattering in the presence of spin-orbit interactions  and noncollinear or dynamic magnetic moments in the barrier. The spin-flip scattering may be included in the tunneling Hamiltonian \eqref{Hprime} as:
\begin{equation}
\mathcal{H}'=\sqrt{\frac{\hbar}{2\pi}}\frac{\tau}{\mathcal{V}}\sum_{\mathbf{k},\mathbf{k}',\sigma, \sigma'}c_{\mathbf{k},\sigma,L}^\dagger c_{\mathbf{k}',\sigma',R} (\delta_{\sigma \sigma'} + \nu\delta_{\bar{\sigma} \sigma'}) + {\rm H.c.}\,,
\label{tunneling}
\end{equation}
where $\bar{\sigma}\equiv-\sigma$ and $\nu$ parametrizes a (random)  spin-flip tunneling amplitude. In perturbation theory, the latter term effectively swaps the spin conserving results for the parallel and antiparallel configurations, so that
\begin{equation}
I\propto\sin^2\frac{\vartheta}{2}+|\nu|^2\cos^2\frac{\vartheta}{2}\,,
\end{equation}
with the proportionality constant obtained from Eq.~\eqref{Iqk} for $\nu=0$.

\section{Berry-phase-induced pumping}
\label{bpp}

\begin{figure}[t]
\includegraphics[width=0.8\linewidth]{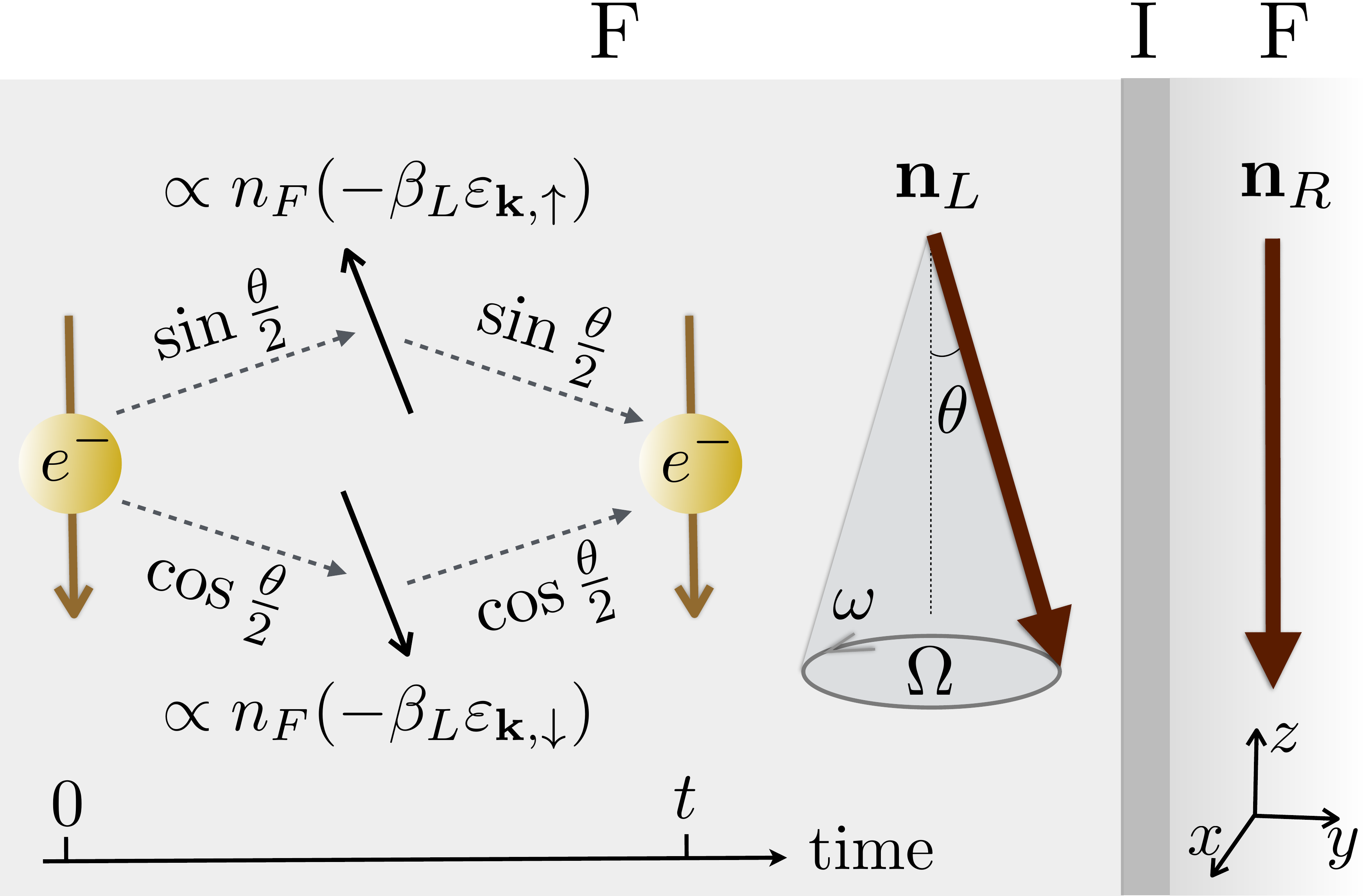}
\caption{A  F$|$I$|$F junction biased by a macrospin precession in the left lead. The two leads have the same volume $\mathcal{V}$: The size of the right lead is expanded only for the sake of illustration. In the left ferromagnet, the order parameter $\mathbf{n}_{L}$ precesses with constant frequency $\omega$ around the $-z$ axis with  cone angle $\theta$. $\mathbf{n}_{R}\to-\mathbf{z}$ in the right ferromagnet is fixed.  $\Omega$ is the solid angle subtended by a cycle of precession of $\mathbf{n}_{L}$. On the left, we sketch the process encoded in the Green function $G^{>}_{\mathbf{k},\downarrow,L}(\varepsilon)$. In the adiabatic limit, a spin-down electron, injected at time $0$ and extracted at time $t$, is a linear combination of  instantaneous spin eigenstates pointing along $\pm\mathbf{n}_{L}$. Both components acquire different phases when following rigidly the precessing  order parameter, leading to probabilities proportional to $\cos^{2}\frac{\theta}{2}$ ($\sin^{2}\frac{\theta}{2}$) and to the number of available states, $1-n_{F}(\beta_L\epsilon_{\mathbf{k},\downarrow(\uparrow)})\equiv n_{F}(-\beta_L\epsilon_{\mathbf{k},\downarrow(\uparrow)})$.}
\label{figure3}
\end{figure}

The connection between the semiclassical pumping by magnetization dynamics discussed in Sec.~\ref{scp} and the quantum-kinetic description of Sec.~\ref{qkt} is revealed by a  field-theoretic tunneling treatment to a classically precessing magnetic bilayer, as sketched in Fig.~\ref{figure3}. The lattice and electronic structures are assumed to be mirror symmetric  as before. When the spin density order parameter $\mathbf{n}_{L}$ precesses steadily with cone angle $\theta$, the left lead is  out of thermodynamic equilibrium and the  theory used previously is not applicable anymore.  The non-equilibrium Keldysh Green function formalism is suited to handle such situation. In terms of the lesser and greater Green functions the tunneling current reads ~\cite{bruusBOOK04, Note4}
\begin{align}
I=\frac{e |\tau|^{2}}{2\pi \mathcal{V}^{2}}
 \int_{-\infty}^{\infty} \frac{d\varepsilon}{2\pi}
 \sum_{\mathbf{k}, \mathbf{k}',\sigma} \big[  &G^{>}_{\mathbf{k},\sigma,L}(\varepsilon) G^{<}_{\mathbf{k}',\sigma,R}(\varepsilon) \nonumber \\
&-G^{<}_{\mathbf{k},\sigma,L}(\varepsilon) G^{>}_{\mathbf{k}',\sigma,R}(\varepsilon) \big]\,.
\label{currentgreeaterlesser}
\end{align}

We chose a conventional basis set in  which the Dirac string is oriented along the negative \textit{z} axis (i.e., through the south pole). In the adiabatic limit, $\hbar\omega\ll\Delta$, the Green functions can be  evaluated  as illustrated in Fig.~\ref{figure3}, keeping track of both the dynamic and geometric (i.e., Berry) phases acquired by the electrons in the precessing exchange field. The former are governed by the instantaneous energies of the electronic states $|\mathbf{k}, \mp \mathbf{n}_{i}\rangle$ (with $i=R,L$), which are given by $\varepsilon_{\mathbf{k},\uparrow ( \downarrow)}= \varepsilon_{\mathbf{k}}\pm\Delta/2$. The geometric phases correspond to the solid angle spanned by the spin trajectories.\cite{berryPRSLA84} The spin-down greater Green function then reads 
\begin{align}
\frac{i G^{>}_{\mathbf{k},\downarrow,L}(\varepsilon)}{2\pi}&=\cos^{2} \frac{\theta}{2}\,n_{F}(-\beta_L\varepsilon_{\mathbf{k,\downarrow}})\delta ( \varepsilon - \varepsilon_{\mathbf{k},\downarrow} + \gamma_-) \nonumber \\
&+\sin^{2}  \frac{\theta}{2}\,n_{F}(-\beta_L\varepsilon_{\mathbf{k, \uparrow}})\delta ( \varepsilon - \varepsilon_{\mathbf{k},\uparrow} + \gamma_+)\,,
\label{eq37}
\end{align}
where $\gamma_{\pm}= \hbar \omega (1 \pm \cos \theta)/2$ accounts for the Berry-phase contribution to the electron phase following adiabatically the precession of the order parameter $\mp\mathbf{n}_{L}$. This phase is, per cycle of precession, $1/2$ (i.e., electron spin) times the enclosed solid angle. Equation~(\ref{eq37})  has the geometric interpretation  shown in Fig.~\ref{figure3}. Along the same lines, the spin-up greater Green function reads
\begin{align}
\frac{i G^{>}_{\mathbf{k},\uparrow,L}(\varepsilon)}{2\pi}=&\cos^{2}  \frac{\theta}{2} \,n_{F}(-\beta_L\varepsilon_{\mathbf{k, \uparrow}})  \delta ( \varepsilon - \varepsilon_{\mathbf{k},\uparrow}-\gamma_-) \nonumber \\
+ &\sin^{2} \frac{\theta}{2} \,n_{F}(-\beta_L\varepsilon_{\mathbf{k, \downarrow}}) \delta ( \varepsilon - \varepsilon_{\mathbf{k},\downarrow}-\gamma_+) \,.
\label{eq38}
\end{align}
The spin-up and spin-down lesser Green functions of the left lead are given by similar expressions, with the replacement of $n_F(-\beta\varepsilon)\to n_F(\beta\varepsilon)$.

The Green functions of the right lead are recovered by simply setting $\theta\to0$. Thereby, Eq.~(\ref{currentgreeaterlesser}) reduces to the simple form
\begin{equation}
I_m= -|\tau|^{2}\frac{e\hbar\omega}{4}(D^{2}_{\uparrow} - D^{2}_{\downarrow})\sin^{2} \theta\,,
\end{equation}
which, using  Eq.~\eqref{equation6},  reproduces Eq.~(\ref{equation2}). The structure of the above Green functions precisely mimics the rotating-frame analysis of Ref.~[\onlinecite{tserkovPRB08tb}].

Several observations link the quantum-kinetic and semiclassical considerations. First of all, we see that the introduction of the Berry phases through the $\gamma_\pm$ energy shifts in Eqs.~\eqref{eq37} and \eqref{eq38} allows us to identify the first two terms on the right-hand sides of  Eqs.~(\ref{epsilontilde}) and~(\ref{equazione26}) as the dynamic and geometric contributions to the spin-$\sigma$ electron energies, respectively. Indeed, writing 
\begin{equation}
\frac{n_B}{2S}=\frac{1-\cos\theta}{2}=\sin^2\frac{\theta}{2}\,,
\label{nB}
\end{equation}
where $n_B$ is now the (average) number of magnons corresponding to the coherent precession with angle $\theta$, we see that $\gamma_\sigma=\hbar\omega(\delta_{\sigma,+}-\sigma n_B/2S)$. Equation~\eqref{nB}, furthermore, allows us to geometrically interpret the delta-function weights in spectral functions \eqref{Adown} and \eqref{Aup}. Namely, 
\begin{align}
\frac{A_{\mathbf{k},\downarrow,L}(\varepsilon)}{2\pi}=&\frac{i}{2\pi}\left[G_{\mathbf{k},\downarrow,L}^>(\varepsilon)-G_{\mathbf{k},\downarrow,L}^<(\varepsilon)\right]\nonumber\\
=&\left(1-\frac{n_B}{2S}\right)\delta ( \varepsilon - \varepsilon_{\mathbf{k},\downarrow}+\gamma_-)\nonumber\\
&+\frac{n_B}{2S}\delta (\varepsilon-\varepsilon_{\mathbf{k},\uparrow}+\gamma_+)\,,
\label{AB}
\end{align}
by inserting Eq.~\eqref{nB}. The semiclassical treatment of this section, however, requires that $n_B\gg1$, hence the fermionic factors $\propto n_F\ll n_B$ are, unfortunately,  not captured in this approximation.\cite{Note5}

We thus conclude that the semiclassical treatment of the magnonic thermal pumping leading to the results of Sec.~\ref{scp}, particularly Eq.~\eqref{equation14}, captures only  part of the story. Namely, these results correspond to a hypothetical situation in which the magnons experience a thermal bias while the electrons are thermally equilibrated. It is more physical to suppose that electrons and magnons are in a common thermodynamic equilibrium in each lead (assuming the electron-magnon coupling in the leads is stronger than the phonon-mediated magnon-magnon coupling across the barrier). This is accounted for by resorting to a more thorough quantum-kinetic description of Sec.~\ref{qkt}, which captures magnonic and electronic spin fluctuations (and the associated pumping and torques) on equal footing.

\section{Discussion and outlook}

We analyzed the thermopower in magnetic tunnel junctions, focusing on the role of collective (transverse) spin fluctuations and the associated magnon-assisted electron transport. We developed a semiclassical framework for calculating stochastic pumping of charge by the magnetic thermal noise as well as a more complete quantum-kinetic theory that captures  the effects of magnonic and electronic noise. The Berry phase accumulated by electrons in the presence of  magnetic fluctuations unifies the two approaches. 

The electronic contribution to the Seebeck effect, $\mathcal{S}_{f} \sim(k_B/e)T/T_{F}$, is found to be augmented by a magnon-assisted tunneling charge current with thermopower $\mathcal{S} \sim(k_B/e)(T/T_{C})^{3/2}$. Focusing on elemental transition metals and taking bulk parameters for $T_F$ and $T_C$ from Refs.~\onlinecite{FermiT} and \onlinecite{Kittel}, respectively, the room-temperature magnonic contribution $\mathcal{S}$ to the thermopower of symmetric tunnel junctions is obtained then to be significant when compared with the free-electron one. Similar estimates have been made in the context of the measured bulk thermopower in metallic ferromagnets.\cite{watzmanPRB16}

We note, however, that the crude $\propto T/T_F$ estimate for the electronic thermopower neglects the nontrivial band-structure features (which enhance the electron-hole asymmetry) in the tunneling density of states, depending on disorder and alloy scattering.\cite{CzernerPRB2011} (In this regard, we point out that also the magnonic band structure may be engineered or tuned to enhance the magnon-assisted thermopower.) It is worthwhile noting that the Curie temperature $T_C$ of transition metals is systematically higher than $T_c\equiv A(\hbar/s)^{1/3}/k_B$ that enters in our expression for the thermopower \eqref{Sfinal} (see, e.g., Ref.~\onlinecite{shiranePRL65} for Fe). We may, furthermore, expect for the magnonic contribution to be enhanced in thin magnetic films forming tunnel junctions, which have larger thermal fluctuations (as manifested by lower Curie temperatures and suppressed long-range order). In light of these physical uncertainties, we are omitting model-dependent numerical factors in the estimates for $\mathcal{S}_f$ and $\mathcal{S}$ (e.g., $\sim\pi^2/3$ and $\pi^{-2}$, respectively, for parabolic bands and featureless tunneling matrix elements). The only certain conclusion, at this point, is that the magnonic contribution to the thermopower can generally not be neglected and may even dominate the thermoelectric properties, especially in the structures of reduced dimensions.

Whereas the order of magnitude of the observed thermopower\cite{walterNATM11,liebingPRL11,linNATC11} of magnetic tunnel junctions is in line with the theoretical estimates,  a quantitative and material-dependent comparison of theory and experiments is difficult also due to the experimental uncertainties. Further measurements of the junction thermopower, as a function of temperature and magnetic field, are called for. Future theoretical work should address the role of phonons, multi-magnon scattering processes, and the enhanced magnetic fluctuations  in the vicinity of $T_C$. While we focused on the mirror-symmetric F$|$I$|$F junctions at low temperatures, it should be interesting to study asymmetric junctions as well as effects stemming from spin relaxation and nonadiabaticity of electron spin dynamics (which is especially relevant when approaching $T_C$), from both quantum-kinetic and semiclassical perspectives.

In the broader terms, the capacity of low-energy collective modes (here, spin waves) to disrupt the electron particle-hole symmetry on energy scales much lower than $T_F$ open new strategies for enhancing thermoelectric characteristics. An interesting open issue concerns the effect of magnons and other collective modes on the large thermopower predicted in heterostructures with magnetic and superconducting orderings.\cite{MachonPRL2013}

\acknowledgments

This work is supported by the ARO under Contract No.~W911NF-14-1-0016,  by the European Research Council, the D-ITP consortium, a program of the Netherlands Organization for Scientific Research (NWO) that is funded by the Dutch Ministry of Education, Culture, and Science (OCW), and by Grants-in-Aid for Scientific Research (Grant Nos.~25247056, 25220910, 26103006)

\end{document}